\documentclass[aps,prb,twocolumn,superscriptaddress,floatfix]{revtex4}
\usepackage{epsfig,amsmath,amssymb,color}
\begin{document}

\title{Bond-impurity induced bound states in disordered  spin-$1/2$  ladders}

\author{M. Arlego}
\affiliation{Departamento de F\'{\i}sica, Universidad Nacional de La Plata,
  C.C.\ 67, (1900) La Plata, Argentina}
\affiliation{Facultad de Ingenier\'{\i}a, Universidad de Lomas de Zamora,
  Camino de Cintura y Juan XXIII, (1832) Lomas de Zamora, Argentina}
\author{W. Brenig}  
\affiliation{Technische Universit\"at Braunschweig, Institut f\"ur Theoretische Physik, Mendelssohnstrasse 3, 38106 Braunschweig, Germany}
\author{D. C. Cabra}
\affiliation{Universit\'{e} Louis Pasteur, Laboratoire de Physique Th\'{e}orique, 3 Rue de l'Universit{\'e},
67084 Strasbourg Cedex, France}
\author{F. Heidrich-Meisner}      
\affiliation{Technische Universit\"at Braunschweig, Institut f\"ur Theoretische
  Physik, Mendelssohnstrasse 3, 38106 Braunschweig, Germany}
\author{A. Honecker}
\affiliation{Technische Universit\"at Braunschweig, Institut f\"ur Theoretische
  Physik, Mendelssohnstrasse 3, 38106 Braunschweig, Germany} 
\affiliation{Universit\"at Hannover, Institut f\"ur Theoretische
  Physik, Appelstrasse 2, 30167 Hannover, Germany}
\author{G. Rossini}
\affiliation{Departamento de F\'{\i}sica, Universidad Nacional de La Plata,
  C.C.\ 67, (1900) La Plata, Argentina}  
\affiliation{Facultad de Ingenier\'{\i}a, Universidad de Lomas de Zamora,
  Camino de Cintura y Juan XXIII, (1832) Lomas de Zamora, Argentina}

\date{May 11, 2004}
\begin{abstract}
We discuss the effect of weak  bond-disorder in two-leg spin ladders on the
dispersion relation of the elementary triplet excitations with a particular focus on the
appearance of bound states in the spin gap. Both the cases of
modified exchange couplings on  the rungs  and the legs of the ladder are analyzed.
Based on a projection on the single-triplet subspace, the single-impurity and small cluster
problems
are treated analytically in the  strong-coupling limit. Numerically, we study the problem 
of a single impurity in a spin ladder 
by exact diagonalization  to obtain
the low lying excitations.
At finite concentrations and to leading order in the inter-rung coupling, we compare the spectra obtained
from numerical diagonalization of large systems within the single-triplet subspace with the results of diagrammatic techniques,
namely low-concentration and coherent-potential approximations. The contribution of  small impurity 
clusters to the density of states is also discussed.
\end{abstract}

\maketitle
\section{Introduction}
\label{sec:1}
Since the synthesis of spin ladder materials 
such as SrCu$_2$O$_3$ (Ref.~\onlinecite{hiroi91})
or (Sr,Ca,La)$_{14}$Cu$_{24}$O$_{41}$ (Ref.~\onlinecite{carron88}) 
and the later discovery of superconductivity
under high pressure in Sr$_{0.4}$Ca$_{13.6}$Cu$_{24}$O$_{41.84}$ (Ref.~\onlinecite{uehara94}),
 spin-$1/2$ two-leg spin ladders have been on the focus of theoretical 
activities (see Refs.~\onlinecite{dagotto96} and \onlinecite{dagotto99} for  review).
This model  is of particular interest, as the ground-state is nonmagnetic and
it is an example for a spin liquid. 
Many properties of the pure system, such as the dispersion of elementary
excitations\cite{barnes93,reigrotzki94,greven96,weihong98}
and the thermodynamics\cite{johnston00} are well understood. The elementary excitations are propagating, massive triplet-modes.
\\
\indent
A natural extension of the   pure spin model
comprises the inclusion of impurities. In this paper,
we discuss the effect of impurities on the dispersion relation of the elementary triplet excitations and focus
on the appearance of bound states in the spin gap. Such states could be visible in resonant  experiments.
One may distinguish between different kinds of impurities. First, magnetic ions, such as the Cu$^{2+}$ ions in 
SrCu$_2$O$_3$,
can be replaced by nonmagnetic ones such as Zn (see Ref.~\onlinecite{sigrist96} and references therein), effectively removing a spin-$1/2$ moment, 
or by other ions with the  same or a different effective moment.
The replacement of the spin-carrying ion  will be referred to as a site impurity. Note that it is
also conceivable that a site impurity leads to modifications of the exchange couplings to neighboring sites.
Second, and this is what we mainly have in mind in this study, the exchange paths 
themselves can be modified by doping the  bridging X-ions
in, e.g., Cu-X-Cu bonds, realizing what we call a bond impurity (or simply impurity) in the following.  
Such a situation is described in Ref.~\onlinecite{manaka01}  for the 
alternating spin chain system (CH$_3$)$_2$CHNH$_3$Cu(Cl$_x$Br$_{1-x}$)$_{3}$
where Cl and Br ions are substituted with each other. Furthermore,   a spin ladder material exists, namely
(C$_5$H$_{12}$N)$_2$CuBr$_4$ (Ref.~\onlinecite{watson01}) where one could think of analogous doping experiments.
This  material   is suggested to contain 
two-leg spin ladders in the strong-coupling limit, i.e.,
the coupling constant along the legs $J_{\mathrm{L}}$  is small compared to 
the coupling along the rungs $J_{\mathrm{R}}$; see Ref.~\onlinecite{watson01}. Moreover, there are  a number of further candidates 
for organic spin ladder materials in the strong-coupling limit, see, e.g., Ref.~\onlinecite{landee01}.
As an  example for an inorganic system,
we mention CaV$_2$O$_5$  for which  a ratio of  $J_{\mathrm{L}}/J_{\mathrm{R}}\sim 0.1$ 
is discussed\cite{konst00}.  
\\
\indent
In the literature, bond randomness in spin ladder systems has been studied both in the weak and strong disorder limit
using the real-space renormalization group method\cite{melin02,yusuf02}, bosonization\cite{orignac98}, and a mapping on 
random-mass Dirac fermions\cite{steiner98}.  Most of these studies have focused on the stability of the ground state
and the gap against disorder
and they find that disordered spin ladders exhibit nonuniversal thermodynamic properties (see, e.g., Ref.~\onlinecite{yusuf02})
 similar to disordered dimerized spin-$1/2$ chains\cite{hyman96}.\\
\indent 
The plan  of the   paper is the following. First, we introduce the model and perform a projection on the one 
triplet subspace in Sec.~\ref{sec:2}. This approximation provides results which are correct in leading order of the inter-rung 
coupling and are quantitatively relevant for spin ladder materials in the strong-coupling limit. 
Second, the single-impurity problem is solved analytically in Sec.~\ref{sec:3} in 
the strong-coupling limit. Also, we analyze small clusters of bond impurities 
on neighboring bonds. The results, i.e., the eigenenergies of single-impurity (anti-) bound states are then compared to 
those of a Lanczos study  for the full spin ladder model. 
In Sec.~\ref{sec:4}, finite concentrations of bond-impurities are considered in the strong-coupling limit.
This problem is tackled both numerically and analytically by means of  diagrammatic approaches  [low-concentration
 approximation and coherent-potential approximation (CPA)]. The comparison of the   numerical results with 
the analytical approaches provides insight into the validity of the latter methods.
Finally, our conclusions are summarized in Sec.~\ref{sec:5}

\section{Model}
\label{sec:2}

The Hamiltonian of the pure two-leg spin ladder reads
\begin{equation}  
        H_0 = \sum_{l=1}^{N}\lbrack J_{\mathrm{R}}\vec{S}_{l,1}\cdot \vec{S}_{l,2}
	 + J_{\mathrm{L}}(\vec{S}_{l,1}\cdot
              \vec{S}_{l+1,1}+ \vec{S}_{l,2}\cdot \vec{S}_{l+1,2})\rbrack  .\label{eq:m1} 
 \end{equation}
 $\vec{S}_{l,1(2)}$ are spin-$1/2$ operators acting on site $l$ on  leg $1(2)$ and $N$ is the number of rungs.
 For the remainder of the paper, we set $J_{\mathrm{R}}=1$ in all explicit computations, but we keep $J_{\mathrm{R}}$ 
 in the equations for clarity.
We will discuss the following situations

\begin{equation}
H=H_0+H'; \quad H'=\sum_{n=1}^{N_{\mathrm{imp}}} h_n
\label{eq:H}
\end{equation}
where $N_{\mathrm{imp}}$ is the number of modified couplings and $h_n$ is
the local perturbation at site $l_n$   leading to either a modified on-site rung interaction
$J_{\mathrm{R}}'=J_{\mathrm{R}}+\delta J_{\mathrm{R}}$ or a modified leg coupling
$J_{\mathrm{L}}'=J_{\mathrm{L}}+\delta J_{\mathrm{L}}$, connecting sites $l_n$ and $l_n+1$.
Explicitly, $h_n$ reads
\begin{eqnarray}
h_n&=& \delta J_{\mathrm{R}}\, \vec{S}_{l_n,1}\cdot
\vec{S}_{l_n,2}\label{eq:m2} \\ 
h_n&=& \delta J_{\mathrm{L}}\,
\vec{S}_{l_n,j}\cdot \vec{S}_{l_{n+1},j}; \quad j=1,2\label{eq:m3} .
\end{eqnarray}
The effect of modified interactions on the one-triplet dispersion will be discussed
in the strong-coupling limit  $J_{\mathrm{L}}\ll J_{\mathrm{R}}$ by projecting on the 
one-triplet subspace. Therefore, all terms contained in $H_0$ destroying or creating  two triplet excitations 
are neglected.
For the latter application of diagrammatic techniques, it is useful to map the spin operators on so called 
bond-operators\cite{chubukov89,sachdev90}
$s_l^{(\dagger)}, t_{\alpha,l}^{(\dagger)};$ $\alpha=x,y,z$. $s_l^{\dagger}$ creates a singlet on 
the $l$th rung out of the vacuum state $|\,0\rangle$ and 
$ t_{\alpha,l}^{\dagger}$ creates a triplet excitation with orientation $\alpha$, respectively.
The exact representation of $S^{\alpha}_{l,1(2)},
~\alpha=x,y,z$, in terms of  bond-operators reads\cite{sachdev90}
\begin{equation}
S^{\alpha}_{l,j}= (1/2)\lbrace\pm   s^{\dagger}_l t_{\alpha,l}^{}\pm t^{\dagger}_{\alpha,l} s_{l}^{ }
-i\epsilon_{\alpha\beta\gamma} t^{\dagger}_{\beta,l} t^{}_{\gamma,l}\rbrace\, .
\label{eq:m4}
\end{equation}
The plus sign corresponds to $j=1$ and the minus sign to $j=2$; 
   $j$ labeling the leg.
To avoid unphysical double occupancies one has to impose the local constraint
(summation over repeated indices is implied in the following)
\begin{equation}
s^{\dagger}_ls^{ }_l+ t_{\alpha,l}^{\dagger}t_{\alpha,l}^{ }=1.
\label{eq:5}
\end{equation}
Projecting on the one-triplet subspace and thereby applying a Holstein-Primakoff type of 
approximation\cite{chubukov89,starykh96} $s_l^{ }=s_l^{\dagger}\approx 1$
results in 
the effective Hamiltonian 
 \begin{equation}
H_{0,\mathrm{eff}}=J_{\mathrm{R}} \sum_l  t^{\dagger}_{\alpha,l}t_{\alpha,l}^{ }  
+\frac{J_{\mathrm{L}}}{2}\sum_l  ( t^{\dagger}_{\alpha,l+1} t_{\alpha,l}^{ } + \mbox{H.c.} )
\label{eq:m6} 
\end{equation}
where we have dropped irrelevant additive constants. 
$H_{0,\mathrm{eff}}$ is diagonalized by a  Fourier transformation 
$t_{\alpha,l}^{\dagger}= (1/\sqrt{N})\sum_k e^{-i k l} t_{\alpha,l}^{\dagger}$ 
leading to
\begin{equation}
H_{0,\mathrm{eff}} = \sum_{k} \epsilon_k t^{\dagger}_{\alpha,k} t_{\alpha,k}^{ } 
\label{eq:m7}\end{equation}
and the dispersion relation of one-triplet excitations is\cite{barnes93}
\begin{equation}
\epsilon_k= J_{\mathrm{R}}+ J_{\mathrm{L}}\cos(k) . \label{eq:m8}
\end{equation}
The perturbations $h_n$  caused by modifications of the exchange couplings  are expressed in terms 
of bond-operators as follows
\begin{equation} 
h_n = \frac{1}{N}\sum_{k,k_1} v_{\mathrm{R(L)}}(k,k_1)\,t_{\alpha,k}^{\dagger} t_{\alpha,k_1}^{ } \label{eq:m9}
\end{equation}
with the potentials $v_{\mathrm{R(L)}}(k,k_1)$ given by
\begin{eqnarray}
v_{\mathrm{R}}(k,k_1) & =& \delta J_{\mathrm{R}}\, e^{i l_n \Delta k}\label{eq:m10}\\
v_{\mathrm{L}}(k,k_1) & =& \frac{\delta J_{\mathrm{L}}}{4}(e^{i l_n \Delta k} e^{ik_1}+e^{-i l_n \Delta k}e^{-ik}) \label{eq:m11}
\end{eqnarray}
where $\Delta k= k_1-k$ is the momentum transferred in a scattering process.
All together, the effective Hamiltonian takes the form 
\begin{equation}
H_{\mathrm{eff}}=H_{\mathrm{0,eff}}+H'_{\mathrm{eff}};\quad
H'_{\mathrm{eff}}=\sum_{n=1}^{N_{\mathrm{imp}}} h_n \label{eq:m12}.
\end{equation}

\section{The single-impurity problem and small impurity clusters}
\label{sec:3}
The solution of the one-impurity problem for the effective Hamiltonian Eq.~(\ref{eq:m12}), i.e., one modified coupling, is derived from 
 Schr\"odinger's equation in real space.  Here, we briefly
outline the procedure and our results, referring the reader to the literature\cite{white}
for details. In addition, small impurity clusters  are addressed and we study the problem of one impurity 
in the full model Eq.~(\ref{eq:H}) using the Lanczos method.

\subsection{Single-impurity problem in real space}
\label{sec:3a}
Schr\"odinger's equation can be cast in the  form
\begin{equation}
\left[ I-G^{0}(E) \, H'_{\mathrm{eff}}\right] | \, \psi \rangle=0,
\label{eq:sin1}
\end{equation}
where $G^{0}(E)=\left( E-H_{0,\mathrm{eff}}\right) ^{-1}$ is the free  Green's function
operator associated to the effective one-particle Hamiltonian $H_{0,\mathrm{eff}}$ in
Eq.~(\ref{eq:m6})  and $| \, \psi \rangle$ is an eigenstate of the Hamiltonian $H_{\mathrm{eff}}$.
A real-space representation can be given using the one-triplet basis
$t_{\alpha,l}$, where  matrix elements of $G^0(E)$ are diagonal in $\alpha$ 
and depend only
on the distance $\Delta l=|l-l'|$. In the continuum limit $N \to \infty$,
they are given by
\begin{equation}
\left[ G^{0} (E) \right]_{\Delta l}^{\alpha ,\beta}=\delta _{\alpha \beta }
\frac{1}{2\pi }\int\limits_{-\pi }^{\pi }\frac{
\cos(k \Delta l)}{E-J_{R} -J_{L}\cos\ k}\ dk \ ,
\label{eq:sin2}
\end{equation}
with $\alpha,\beta=x,y,z$.
Although $E$ needs to be analytically continued to the complex plane in 
order to obtain the retarded Green's function by setting $E\to E+i0^+$, notice that for real $E$, $ 
\vert E-J_{\mathrm{R}} \vert >J_\mathrm{L}$, matrix elements in Eq.~(\ref{eq:sin2}) are real.
For simplicity, we place the modified rung-coupling  on site $l=0$ and the modified
leg-coupling between sites $l=0$ and $l=1$. The (anti-) bound states are found by setting the
determinant of $I-G^{0}(E) \,H'_{\mathrm{eff}}$ to zero, and, due to the impurity location,
at most the upper $2\times 2$ submatrix needs to be considered. Notice, however, that the
eigenvalues are threefold degenerate because of the three triplet modes $\alpha=x,y,z$.
\\\indent
The eigenenergy of the (anti-) bound state in the single rung-impurity case
$\delta J_R \not= 0,\delta J_L = 0$ is obtained as
\begin{equation}
E_{\mathrm{1,R}} = J_{\mathrm{R}} \pm
\sqrt{J_{\mathrm{L}}^{2}+(\delta J_{\mathrm{R}})^{2}}; \quad\delta J_{\mathrm{R}}\gtrless 0.
  \label{eq:sin3}
\end{equation}
The plus(minus) sign in Eq.~(\ref{eq:sin3}) corresponds to
$\delta J_{\mathrm{R}}>0$ ~ $(\delta J_{\mathrm{R}}<0)$. Therefore,
a bound state in the spin gap, i.e., below the original one-triplet band,
appears for $\delta J_{\mathrm{R}}<0$. Conversely, for $\delta
J_{\mathrm{R}}>0$, there is an anti-bound state above the
one-triplet band.\\
\indent
Analogously, one finds the eigenenergies of
(anti-) bound states in the single leg-impurity case for $\delta J_R =
0,\delta J_L \not= 0$. For $\delta J_{\mathrm{L}}>0$, there are always both a bound and an anti-bound
state, their energies given by
\begin{equation}
E_{\mathrm{1,L}} =J_{\mathrm{R}} \pm \frac{J_{\mathrm{L}}+
\frac{\delta J_{\mathrm{L}}}{2}(1+\frac{\delta J_{\mathrm{L}}}{4J_{\mathrm{L}} })}{
1+\frac{\delta J_{\mathrm{L}}}{2 J_{\mathrm{L}}}};\quad \delta
J_{\mathrm{L}}>0.
\label{eq:sin4}
\end{equation}
On the other hand, we note that there are no states outside the
one-triplet band for $-4 J_{\mathrm{L}}  <\delta
J_{\mathrm{L}}<0$; instead, we expect the appearance of resonant modes
inside the band [see Sec.~\ref{sec:4.2}, Fig.~\ref{fig:5}~(c)].
Finally, for strong
ferromagnetic coupling $\delta J_L < -4 J_L$, Eq.~(\ref{eq:sin4})
again has two solutions;   however, we will
restrict the discussion to the case of antiferromagnetic couplings.
\\
\indent
The wave function for (anti-) bound states $\psi_\alpha(l)$ can also be derived in a closed form:
\begin{eqnarray}
\delta J_{\mathrm{R}}\not= 0 &;& \delta J_{\mathrm{L}}= 0:\nonumber\\
& & \psi_\alpha(l) \propto \lbrack G^0(E_{\mathrm{1,R}})\rbrack_{l}^{\alpha\alpha}\quad (l>0), \label{eq:sin4a}\\
\delta J_{\mathrm{R}} = 0 &;& \delta
J_{\mathrm{L}}\not= 0:  \nonumber\\
& & \psi_\alpha(l) \propto \lbrack G^0(E_{\mathrm{1,L}})
\rbrack_{l}^{\alpha\alpha} +\lbrack G^0(E_{\mathrm{1,L}})
\rbrack_{l-1}^{\alpha\alpha} \nonumber\\
& & \quad (l>1).\label{eq:sin4b} 
\end{eqnarray}
The width of $\psi_\alpha(l)$ in real space 
only depends on the ratio of $\delta J_{\mathrm{R}}/J_{\mathrm{L}}$
(or $\delta J_{\mathrm{L}}/J_{\mathrm{L}}$, respectively). 
The spatial extent of 
$|\psi_\alpha(l)|^2$ is the narrower, the larger this ratio is. 
For instance, $|\psi_\alpha(l=4)|^2/|\psi_\alpha(l=1)|^2 < 0.01 $ for
$\delta J_{\mathrm{R}}/J_{\mathrm{L}}=1, \delta J_{\mathrm{L}}=0$, while 
$|\psi_\alpha(l=4)|^2/|\psi_\alpha(l=1)|^2 \sim 0.15 $ for 
$\delta J_{\mathrm{R}}/J_{\mathrm{L}}=1/3$.

\subsection{Impurity clusters}
\label{sec:3b}
 Solving Schr\"odinger's equation Eq.~(\ref{eq:sin1}) in real space 
allows for the discussion of small impurity clusters. We consider  the presence of
impurities of the same type located on some of the first $N_c$ ladder sites. The
(anti-) bound eigenenergies depend in principle on both the number of modified couplings
and their distance as well as the perturbation  $\delta J_{\mathrm{R[L]}}$ itself.
It is natural to expect the one-impurity eigenstates to interfere
when the single impurities come close enough.
\\\indent
As before, the eigenenergies of the cluster are evaluated by setting the determinant of
$I-G^{0}\left(E\right) \,H'_{\mathrm{eff}}$ to zero. Now, only the upper $N_{c}\times
N_{c}$ submatrix needs to be considered. As an example, we give the analytical expression for the
case of two modified rung couplings on neighboring sites ($N_{c}=2$).
One solution exists for both $\delta J_{\mathrm{R}}<0$ (inside the gap) or 
$\delta J_{\mathrm{R}}>0$ (above the one-triplet band). Their eigenenergies 
$E_{2,\mathrm{R}}$ read
\begin{equation}
E_{2,\mathrm{R}} = \frac{J_{\mathrm{L}}J_{\mathrm{R}}\pm J_{\mathrm{L}}^2 + 2
J_{\mathrm{L}}\delta J_{\mathrm{R}} \pm 2 \delta
J_{\mathrm{R}}(J_{\mathrm{R}}+\delta J_{\mathrm{R}}) }{J_{\mathrm{L}}\pm
2\delta J_{\mathrm{R}}}\,.
\label{eq:sin4c}
\end{equation}
The plus sign has to be used for $\delta J_{\mathrm{R}}>0$ and the minus 
sign in the opposite case.
We have computed similar expressions for clusters up to $N_{c}=5$. 
In Sec.~\ref{sec:4.3}, we will show 
that the influence of such clusters explains  the details of the peak structure 
in the density of states obtained by numerical
diagonalization of systems with a  finite impurity concentration.

\subsection{Comparison with exact diagonalization} 

To test the region of validity of the results derived above to
first order in $J_{\mathrm{L}}/J_{\mathrm{R}}$, we now compare them
to numerical results for the full spin ladder Hamiltonian Eq.~(\ref{eq:H})
with one modified rung or leg coupling.

We have exploited $S^z$-conservation, spin-inversion symmetry, reflection
symmetry at the impurity bond and, in the case of a rung impurity,
exchange symmetry of both legs. 
Usually, one uses periodic boundary conditions and exploits translational invariance.
Although the latter is not possible if an impurity is present, we still  
apply periodic boundary conditions along the legs in order to minimize surface effects. 
 Finite-size effects turn out to be smallest for an
even number of rungs. We  therefore concentrate on systems
with $N=4$, $6$, $8$, $10$, $12$, and $14$ rungs. The largest dimension
is slightly above 10 million and occurs for $N=14$ rungs (28 spins) and
one leg impurity (where exchange symmetry of the legs is absent).

\begin{figure}
\epsfig{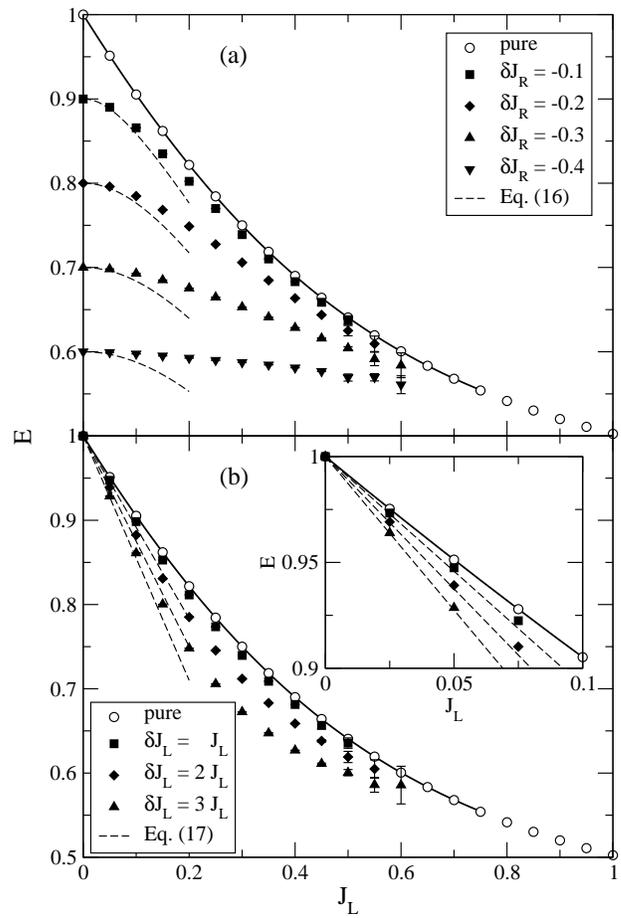}
\caption{
Excitation energy $E$ of the lowest level for the full spin ladder Hamiltonian
(\ref{eq:H}).
Panel (a): one rung impurity
$J_{\mathrm{R}}'=J_{\mathrm{R}}+\delta J_{\mathrm{R}}$.
Panel (b): one leg impurity
$J_{\mathrm{L}}'=J_{\mathrm{L}}+\delta J_{\mathrm{L}}$.
In all cases, the normalization is fixed to $J_{\mathrm{R}}=1$.
Symbols are obtained by extrapolation of Lanczos diagonalization on
finite systems. Open circles are for the pure system
($\delta J_{\mathrm{R}} = 0$ and $\delta J_{\mathrm{L}} = 0$)
and correspond to the spin gap; the solid line is a [7,6] Pad\'e approximant
to the 13th order strong-coupling series\cite{weihong98} for the spin gap of
the pure ladder.
Dashed lines display the analytical result for the position of the
bound state in the effective Hamiltonian Eq.~(\ref{eq:m12}), namely
Eq.~(\ref{eq:sin3}) [panel (a)] and Eq.~(\ref{eq:sin4}) [panel (b)].
}
\label{fig:1}
\end{figure}

In each of the relevant subspaces we have computed the lowest eigenvalue
using the Lanczos procedure. The results for the lowest excitation
energy $E$ at finite $N$ have then been extrapolated to the thermodynamic
limit $N \to \infty$ using the
Vanden-Broeck-Schwartz-algorithm\cite{VBS79,HeSch88}
with $\alpha = -1$. For the pure ladder this yields  estimates for
the spin gap shown by the open circles in Fig.~\ref{fig:1}. Using
$N\le 14$, we find a value of $(0.5025 \pm 0.0008) J_{\mathrm{R}}$
at $J_{\mathrm{L}} = J_{\mathrm{R}}$, in excellent agreement with
accepted values for this case (see section III.A of
Ref.~\onlinecite{johnston00} for a summary). As a further comparison,
the full line in Fig.~\ref{fig:1} shows a [7,6] Pad\'e
approximant to the 13th order series for the spin gap of the pure
ladder of Ref.~\onlinecite{weihong98}.\\
\indent
 A finite number of impurities
(vanishing density) does not affect the one-triplet band in the thermodynamic limit. 
Hence, the
result for the spin gap in the pure case also corresponds to the lower
boundary of the one-triplet band if impurities are present.

Turning now to the case of {\em one} impurity, we concentrate on those
situations where we may expect the lowest excitation to be a bound state
at the impurity, namely $\delta J_{\mathrm{R}} < 0$ or
$\delta J_{\mathrm{L}} > 0$, respectively. 
To understand the finite-size behavior of systems with one impurity, 
it is important to realize that there are now two competing length scales involved.
On the one hand, there is the correlation length of the pure system, and on the 
other hand, the spatial extent of the impurity wave function needs to be considered 
[see Eqs.~(\ref{eq:sin4a}) and (\ref{eq:sin4b})]. Indeed, the typical width of the impurity wave function,
which depends on the actual choice of parameters,  can be (much) bigger
than  the correlation length.
This interplay leads to a crossover in the finite-size behavior.\\\indent
For small $J_{\mathrm{L}}$, the energy of
the impurity level increases with system size which is in contrast
to the behavior of the pure system where the energy of the lower edge of the band decreases with system
size.
Since the latter finite-size behavior is preserved at large
$J_{\mathrm{L}}$, finite-size effects are nonmonotonic in
the intermediate region, i.e., the finite-size behavior changes
at a characteristic system size that increases as
the impurity level approaches the one-triplet band. 
One now has to be more careful
with the extrapolation,
and  we can use only those system sizes which are
in the asymptotic regime for large $N$. Accordingly, Fig.~\ref{fig:1}
shows extrapolated data points for the impurity level only in a
restricted region of $J_{\mathrm{L}}$.
When the impurity level approaches the one-triplet band,
error bars become large, making it difficult to decide whether this level
merges into the band or approaches it only asymptotically. In any case, the
numerical data demonstrate the presence of an impurity bound state in
a wide parameter region.

In the limit of small $J_{\mathrm{L}}\ll J_{\mathrm{R}}$, we can compare
to Eqs.~(\ref{eq:sin3}) and (\ref{eq:sin4}), respectively.
One indeed observes  quantitative agreement for sufficient small
$J_{\mathrm{L}}$, see Fig.~\ref{fig:1}~(a) and  the inset of  Fig.~\ref{fig:1}~(b). 
Note that due to the normalization of Fig.~\ref{fig:1},
i.e., $\delta J_{\mathrm{L}}\sim J_{\mathrm{L}}$, 
Eq.~(\ref{eq:sin4}) results in 
straight lines which start at $E=1$ for $J_{\mathrm{L}}=0$.
At larger $J_{\mathrm{L}}$ deviations can be observed in Fig.~\ref{fig:1}, but
an impurity level can still be seen. Accordingly, the first-order
approximation can still be expected to be qualitatively correct even
in a parameter region where it is no longer quantitatively accurate. Hence,
we may use the first-order approximation to study several impurities and even
finite densities which is no longer systematically possible by Lanczos
diagonalization of the full ladder Hamiltonian.

\section{Finite concentrations}
\label{sec:4}
In this section, we discuss the effect of a finite concentration of modified couplings $J_{\mathrm{R}}$ or $J_{\mathrm{L}}$
on the one-triplet dispersion in the strong-coupling limit $J_{\mathrm{L}}\ll J_{\mathrm{R}}$.
To treat this problem, we apply  diagrammatic techniques, namely, a low-concentration approximation (LCA) 
and the coherent-potential approximation (CPA),
and numerical diagonalization of large systems.
We also use our  analytical results for the eigenenergies of  small impurity clusters to  explain the details 
in the numerical results for the density of states, both for (anti-) bound states and resonance modes.
 \\
 \indent 
Before we turn to the  discussion of  the analytical methods and  compare the results to those from 
numerical impurity averaging (NAV), let us consider certain
limiting cases. In the following, $c$ denotes the concentration of impurities. Note that in the case of
impurities on the legs, we set $c=1$ if all $2N$ couplings are modified. 
\\\indent
The limiting cases are: (i) the pure system ($c=0$); (ii) the  single-impurity case (see Sec.~\ref{sec:3a}); 
(iii)  the case $c=1$ where all couplings are equal to $J_\mathrm{R}+\delta J_\mathrm{R}$ or 
$J_\mathrm{L}+\delta J_\mathrm{L}$, respectively.
In the latter case and for $\delta J_{\mathrm{R}}\not= 0, \delta J_{\mathrm{L}}= 0$, a one-triplet band
with the dispersion $E_k=(J_R+\delta J_R)+J_{\mathrm{L}} \cos(k) $ will result, i.e., its center is shifted 
by $\delta J_R$ with respect to the center of the original band for $c=0$. Therefore, 
the single-impurity (anti-) bound state should  develop into a dispersive band as the concentration increases
while the center $\epsilon(c)$ of the band lies between $J_{\mathrm{R}} - \sqrt{J_{\mathrm{L}}^{2}+(\delta J_{\mathrm{R}})^{2}} < \epsilon(c) <J_\mathrm{R}+\delta J_\mathrm{R}$ 
for $\delta J_{\mathrm{R}}<0$. An analogous scenario arises for $\delta J_{\mathrm{R}}>0$.\\
\indent
For $\delta J_{\mathrm{R}}= 0, \delta J_{\mathrm{L}}> 0$, 
the triplet dispersion in the limit of $c=1$, i.e., all  $J_{\mathrm{L}}$ modified, reads
$E_k= J_{\mathrm{R}}+ (J_{\mathrm{L}}+\delta J_{\mathrm{L}}) \cos(k)$. 
Thus, the bound- and anti-bound states  appear symmetrically with respect to the center of the original band.
On increasing the concentration $c$,
 additional impurity levels will appear and eventually, they will merge in the original band.
 Finally, there will be one broadened band possessing a bandwidth of $(J_{\mathrm{L}}+\delta J_{\mathrm{L}})$.
\begin{figure}
\centerline{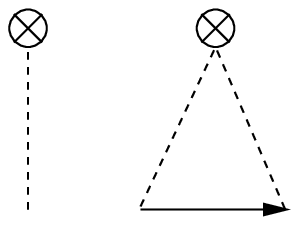}
\caption{Sketch of the diagrammatic expansion of the self energy $\Sigma(E)$ in the low-concentration limit. 
$G_k^0(E)=1/(E-\epsilon_k)$ is the free one-triplet Green's function. }
\label{fig:2}
\end{figure}

\subsection{Low-concentration and coherent-potential approximation}
\begin{figure}
\epsfig{file=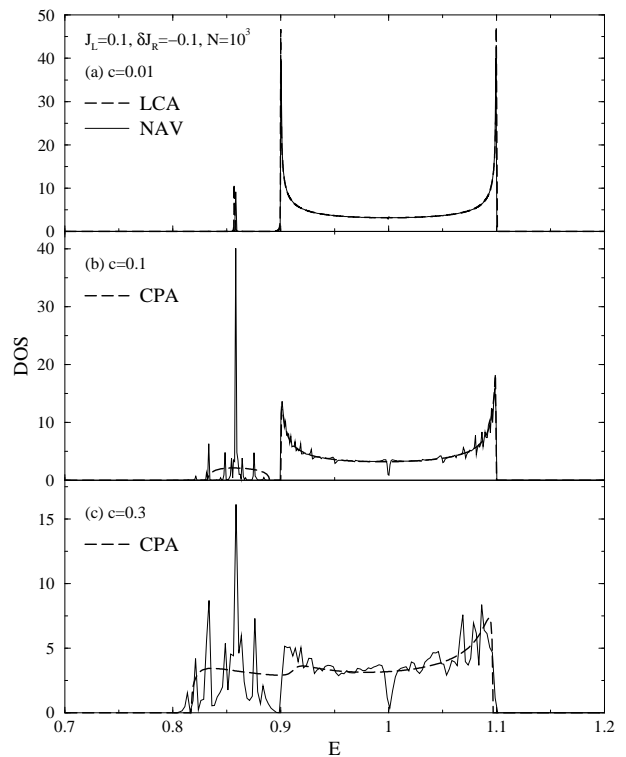,width=0.45\textwidth}

\caption{Density of states (DOS) at finite concentration of modified couplings $J_{\mathrm{R}}'$:
 numerical data (NAV, solid line) for spin ladders with $N=10^3$ rungs 
 and $J_{\mathrm{R}}=1,J_{\mathrm{L}}=0.1,\delta J_{\mathrm{R}}=-0.1$ (concentration:  panel (a) $c=0.01$; panel (b) $c=0.1$; panel (c) $c=0.3$.).
 Dashed line in panel (a): 
 low-concentration approximation (LCA); in panel (b) and (c): CPA.
 }
\label{fig:4}
\end{figure}

Based on a diagrammatic expansion of the one-triplet Green's function in the presence of impurities, 
a number of useful methods exist to get approximate results for the self-energy $\Sigma(E)$.
First, we  briefly comment  on the low-concentration approximation and second, we discuss results from the
 coherent-potential approximation.
\\\indent
As in  Sec.~\ref{sec:3} we will concentrate on
 the  single-triplet subspace.  
Thus, apart from integrating out the singlet, 
the hard-core constraint Eq.~(\ref{eq:5}) is automatically satisfied within our
approximation, i.e., first-order perturbation theory in $J_{\mathrm{L}}/J_{\mathrm{R}}$.  
\\
\\
{\it Low-concentration approximation - }
Using standard impurity-averaging techniques (see, e.g., Ref.~\onlinecite{doniach}), the self-energy of the 
one-triplet Green's function can be obtained in  first order in the impurity concentration.
As the averaging procedure restores translational invariance, the one-triplet Green's function
$G_k(E)$ can be written in terms of the Dyson equation
\begin{equation}
G_k(E)= \frac{1}{E-\epsilon_k -c \Sigma(E)}\,.
\label{eq:imp_av1}
\end{equation}
Keeping only terms linear in  $c$ implies that 
the self-energy $\Sigma(E)$ is equal to the $\hat T$-matrix of the one-impurity problem.
The diagrammatic expansion of the self energy is sketched in Fig.~\ref{fig:2}. 
 Note that all quantities in Eq.~(\ref{eq:imp_av1})  become $2\times 2$ matrices if the modified coupling connects two sites
 as realized by a  leg impurity. For a more detailed discussion of this technique, the reader is referred to, e.g., 
 Ref.~\onlinecite{brenig91}.\\ 
\begin{figure}
\epsfig{file=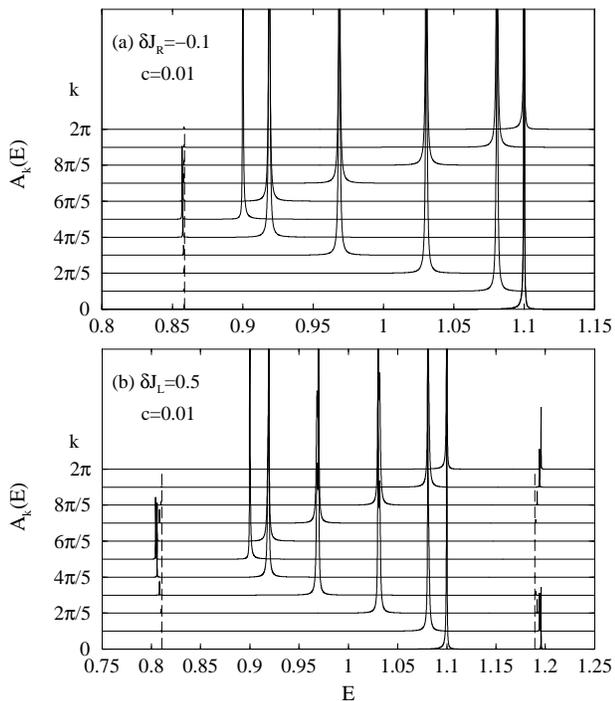,width=0.45\textwidth}
\caption{Spectral function $A_{k}(E)$ in low-concentration approximation (LCA)
 for (a) perturbed rung couplings with $\delta J_{\mathrm{R}}=-0.1$, $c=0.01$; (b) 
 perturbed leg couplings with $\delta J_{\mathrm{L}}=0.5$, $c=0.01$. 
 The dashed lines mark the positions of the (anti-) bound states from Eqs.~(\ref{eq:sin3}) and (\ref{eq:sin4}).
 $J_{\mathrm{R}}=1$ and $J_{\mathrm{L}}=0.1$ in both cases.}
\label{fig:3}
\end{figure}
\indent 
The spectral function $A_{k}(E)=-(1/\pi) \mbox{Im} G_k(E)\,$ is plotted in Fig.~\ref{fig:3} for (a) $J_{\mathrm{L}}=0.1,\delta J_{\mathrm{R}}= - 0.1, c=0.01$ 
and (b) for $J_{\mathrm{L}}=0.1,\delta J_{\mathrm{L}}= 0.5, c=0.01$. 
In accordance with our previous results we find one bound state in case (a) and a bound and an anti-bound state in case (b).
Figure~\ref{fig:3} further reveals that first, the impurity levels have developed a small dispersion and second, 
the spectral weight is concentrated around  $k=\pi$ for the bound states 
while it vanishes in the center of  the zone, and
vice-versa for the anti-bound states.\\\\
{\it Coherent-potential approximation - }
The coherent-potential approximation allows one to interpolate between the two limits of $c=0$ and 
$c=1$. Here, we apply this method to the case of $\delta J_{\mathrm{R}}< 0, \delta J_{\mathrm{L}}= 0$. The self-energy
is obtained from a self-consistent solution of  the equation\cite{elliott74}
\begin{equation}
\Sigma(E) = \frac{c \,\delta J_{\mathrm{R}}}{ 1 -  G(E)\, \lbrack \delta J_{\mathrm{R}}-\Sigma(E)  \rbrack }.
\label{eq:cpa1}
\end{equation}
Rather than deriving this equation (see Ref.~\onlinecite{elliott74} for details), let us mention  some  features of 
this method: (i) the self-energy is symmetric under exchange of 
host and impurity sites, i.e.,
$c$ and $1-c$ and the respective replacement of the coupling constants; and (ii) it gives qualitatively correct results 
for the density of states for intermediate concentrations.
We note that in contrast to the low-concentration approximation [see, e.g., Fig.~\ref{fig:4}~(a)], 
the  CPA does not lead to a sharp peak 
in the density of states at the position of the impurity level  even for low concentrations. This can, for example, be seen
in Fig.~\ref{fig:4}~(b) for $c=0.1$. We have, however, checked that in
both diagrammatic approaches, the total weight  in the impurity levels is the same
and that it
 grows linearly with the impurity concentration, as expected.

\begin{figure}
\epsfig{file=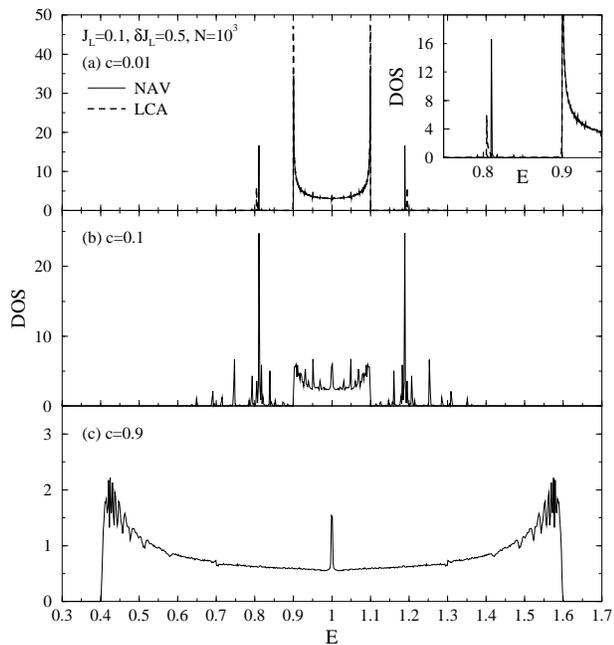,angle=-90,width=0.45\textwidth}
\caption{Density of states (DOS) at finite concentration of modified couplings $J_{\mathrm{L}}'$:
 numerical data (NAV, solid line) for spin ladders with $N=10^3$ rungs 
and $ J_{\mathrm{R}}=1, J_{\mathrm{L}}=0.1,$ and $\delta J_{\mathrm{L}}=0.5$ 
(concentration:  panel (a) $c=0.01$; panel (b) $c=0.1$; panel (c) 
$c=0.9$.). Dashed line in panel (a): low-concentration approximation (LCA).
Inset of panel (a): structure of the DOS in the vicinity of the bound state.
Panel (c): one observes resonance modes inside the band for $c=0.9$; the case 
shown here is equivalent to $J_{\mathrm{L}}= 0.6$, $\delta J_{\mathrm{L}}= -0.5 <0$,  and $c=0.1$.
}
\label{fig:5}
\end{figure}

\subsection{Numerical results and comparison}\label{sec:4.2}
\begin{figure}[t]
\begin{center}
\epsfig{file=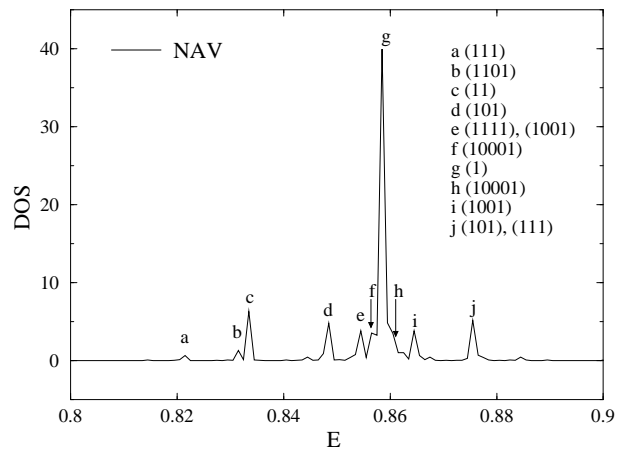,width=0.45\textwidth,angle=0}
\end{center}

\caption{
Comparison of eigenenergies of bound states induced by
 rung-impurity clusters in a clean system,  derived analytically, with numerical 
 diagonalization  of $H_{\mathrm{eff}}$ on large systems
 with a finite concentration of modified rung couplings.
The density of states (DOS) in the vicinity of the one-impurity  level is shown. 
The patterns in parenthesis denote different types of impurity clusters:
with $1$, we indicate the relative position of the rung impurities in the cluster
sequence. The numerical data (NAV, solid line) correspond to $J_{\mathrm{R}}=1,J_{\mathrm{L}}=0.1$,
$\delta J_{\mathrm{R}}=-0.1$ and $c=0.1$. 
The letters 'a' to 'j' relate the peaks in the DOS to certain impurity clusters, which are listed in the legend. }
\label{fig:6}
\end{figure}

Now we compare the analytical results with a numerical diagonalization of the effective Hamiltonian on large systems 
and sampling over several realizations at fixed concentration.
The effective Hamiltonian $H_{\mathrm{eff}}$ Eq.~(\ref{eq:m12}) has been diagonalized  on finite 
systems with  $N=10^3$ rungs  for different choices of 
impurity concentrations $c$ for both types of bond impurities. The density of states (DOS) 
is obtained from  binning the eigenvalues, the  bin-width of typically  $\Delta E \sim 10^{-3} J_{\mathrm{R}}$ determining 
the resolution in Figs.~\ref{fig:4} to \ref{fig:7}. Results are
 shown for $J_{\mathrm{L}}=0.1,\delta J_{\mathrm{R}}=-0.1$  in Fig.~\ref{fig:4}
(panel (a): $c=0.01$; (b): $c=0.1$; (c): $c=0.3$). 
Note that, according to Eq.~(\ref{eq:sin3}), the position of the single-impurity level is $E_{1,\mathrm{R}}=0.8586 J_{\mathrm{R}}$.
The following features are observed:
(i) for increasing concentration, additional peaks appear in the vicinity of
the one-impurity level. They stem from impurity clusters, i.e., impurities occupying neighboring sites, as will be discussed
in more detail below. (ii) The bound state level develops into a band centered around $J_{\mathrm{R}}+\delta J_{\mathrm{R}}=0.9$ 
as a function of concentration $c$. 
Notice that larger concentrations $c>0.5$ are conveniently realized by setting
$c\to 1-c, J_{\mathrm{R}}\to J_{\mathrm{R}}+\delta J_{\mathrm{R}}$, and $\delta
J_{\mathrm{R}}\to -\delta J_{\mathrm{R}}$.
 (iii) Inside the original band, the curve is not smooth, but displays 
small oscillations. These features are neither due to finite-size effects nor due to low statistics (the density of states 
has been obtained by averaging over typically a few thousand random realizations at fixed concentration). 
As we shall discuss below, 
their origin can also be related to the effect
of impurity clusters. \\
\indent
Let us now comment on the comparison of the numerical  with  the analytical results. 
By integrating  $A_{k}(E)$  over the momentum $k$, the density of states $n(E)$ is obtained. Results from the LCA are
compared to the numerical impurity averaging 
 in the case of $J_{\mathrm{L}}=0.1,\delta J_{\mathrm{R}}=-0.1, c=0.01$ in Fig.~\ref{fig:4}~(a).
Both approaches agree well with regard to  
the position of the main impurity level. 
The comparison with the results from the CPA  for $c=0.1$  [Fig.~\ref{fig:4}~(b)]
and $c=0.3$ [Fig.~\ref{fig:4}~(c)] shows that this approach gives qualitatively 
reasonable results even at fairly large concentrations. At $c=0.3$, the impurity levels and the original band
start to merge.
 \\\indent
For a finite concentration of leg couplings, the numerical results confirm our qualitative expectations. 
The data are shown in Fig.~\ref{fig:5} for $J_{\mathrm{L}}=0.1,\delta J_{\mathrm{L}}=0.5$ 
(panel (a): $c=0.01$; (b): $c=0.1$; (c): $c=0.9$). For clarity, we note that the possible
impurity configurations are: (i) one modified coupling on one leg, connecting, e.g., rung $l$ and $l+1$; and (ii)
both couplings between rung $l$ and $l+1$ modified. Both cases are taken into account in the numerical implementation.
\\ \indent
 The impurity levels occur symmetrically with respect to the center of the band. On increasing the concentration
 $c$, the original band widens and eventually includes all impurity levels (see Fig.~\ref{fig:5}~(c) for $c=0.9$).
The influence of impurities is now visible as resonance modes inside the band. Note that the 
last case is equivalent to  $J_{\mathrm{L}}=0.6$, $\delta J_{\mathrm{L}}=-0.5$, and $c=0.1$.\\
\indent
Comparing to the LCA from Eq.~(\ref{eq:imp_av1}) for the case of 
$J_{\mathrm{L}}=0.1,\delta J_{\mathrm{L}}=0.5, c=0.01$ [see Fig.~\ref{fig:5}~(a)],
we see that  
the positions of the highest peaks seen  in the numerical data and the analytical result  almost coincide,
similar to the case shown in Fig.~\ref{fig:4}~(a).
The  inset of Fig.~\ref{fig:5}~(a) contains a zoom of the region around the lower single-leg impurity peak
with $E_{1,\mathrm{L}}= 0.811 J_{\mathrm{R}} $ revealing the presence of  several   less pronounced structures.
\\\indent
In summary, the analytical approaches give fair results for the overall structure of the density of states 
even at large concentrations as exemplified in the case of $\delta J_{\mathrm{R}}< 0, \delta J_{\mathrm{L}}=0$.
The effects of impurity clusters are not taken into account in the diagrammatic description. 

\subsection{Analytical results for small impurity clusters}
\label{sec:4.3}

Next we analyze  the  details of the peak structure of the impurity levels arising from   a
variety of small  impurity clusters for the case of modified rung couplings.
 The clusters analyzed include, e.g.,  the patterns $(11)$, $(101)$, $(111)$, $(1001)$,
$(1101)$, $(1111)$, and $(10001)$ where, in this notation,  $1$
indicates a modified rung in the cluster sequence and $0$ indicates no
impurity placed on a rung site.
For example, $(11)$ denotes two rung impurities on neighboring sites in an otherwise clean system.\\
\indent
 We find that the energy
eigenvalues corresponding to the main  peaks outside the triplet band observed in the numerical
results for the density of states (see  the previous section) can be associated with the contribution  
of certain impurity clusters.
In particular, in  Fig.~\ref{fig:6}, we show the matching between the numerical
bound state structure and the analytically computed
eigenenergies corresponding to different small clusters placed in an otherwise clean system
 for the case of $J_{\mathrm{L}}=0.1$,
$\delta J_{\mathrm{R}}=-0.1$ and $c=0.1$ (see Fig.~\ref{fig:4}~(b) for the full density of states in the same case).

\begin{figure}[t]
 \centerline{\epsfig{file=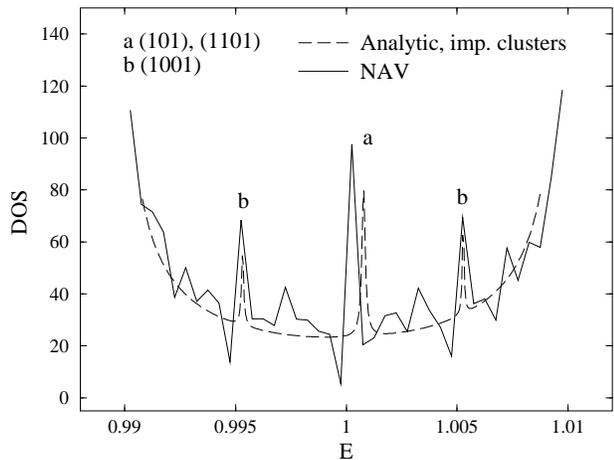,width=0.45\textwidth,angle=0}}
 \caption{Matching between the  peak structure inside the triplet band obtained numerically at finite
  concentration of rung impurities (NAV, solid line) and analytically for systems with different types of 
  impurity clusters (dashed line), for
 $J_{\mathrm{R}} =1$,  $J_{\mathrm{L}} =0.01$,
  $\delta J_{\mathrm{R}}=-0.05$ and $c=0.1$. }
  \label{fig:7}
\end{figure}
Let us now discuss the features inside the triplet band in the presence of
impurities.
The density of states (DOS) $n(E)$ can be evaluated from
\begin{equation}
n(E) = -\frac{1}{\pi }{\mbox{Im}}\,\mbox{Tr}\,G(E)=n_{0}(E)+n_{\mathrm{imp}}(E),
\label{eq:DOS}
\end{equation}
where \ the free density of states\cite{elliott74} is
$n_{0}(E)=-(1/\pi)\, \mbox{Im} \,\mbox{Tr}\,G^{0}(E)$ and the contribution from the impurities $n_{\mathrm{imp}}(E)$
can be written as
\begin{equation}
n_{\mathrm{imp}}(E)=-\frac{1}{\pi } \, \mbox{Im}\frac{d}{dE} \ln \, \lbrack\mathrm{Det}\left( I-G^{0}\left(E\right) \,
H_{\mathrm{eff}}^{\prime }\right) \rbrack.  \label{eq:DOSimp}
\end{equation}

We use Eq.~(\ref{eq:DOSimp}) to compute the
contributions from particular impurity clusters to the DOS inside the triplet band.
Moreover, from the amplitudes of the different peaks outside the one-triplet band, one can read off, at least qualitatively, the
distribution of probabilities for the presence of the different clusters in a random
sample of impurities for a given concentration. This information is in turn  used to
weight the influence of each cluster  on the peak structure inside the triplet band.
In  Fig.~\ref{fig:7}, we show an example where the main peaks
are associated with the corresponding cluster contributions. We
are able to  match, in this particular case,  
the central peak, which is slightly shifted to the right of the band-center,  with
the patterns $(1101)$ and $(101)$, and the two ones,
which are almost symmetrical with respect to the band center,  with a contribution from the $(1001)$ cluster.
\\
\indent 
This analysis explains on the one hand the appearance of the various localized modes and on the other hand 
it gives evidence that  
the peaks inside the triplet band originate from the existence of 
impurity clusters. \\


\section{Discussion and Conclusions}
\label{sec:5}

In this paper we have studied the appearance of bound states in the spin gap of 
spin ladders with bond impurities. Both the cases of modified rung and leg couplings 
have been considered. We have derived analytical results for the position of 
bound states in  the strong-coupling limit equivalent to first order perturbation theory
in $J_{\mathrm{L}}/J_{\mathrm{R}}$. The existence of impurity induced bound states
has been verified by a Lanczos study of finite spin ladders with one impurity
and $0<J_{\mathrm{L}}\leq J_{\mathrm{R}}$ and we find that our analytical results are 
quantitatively correct for $J_{\mathrm{L}}\lesssim  J_{\mathrm{R}}/10 $
and that a qualitative agreement is still found  for larger $J_{\mathrm{L}}$. Recently discovered spin ladder
materials such as,
for example, 
(C$_5$H$_{12}$N)$_2$CuBr$_4$ (Ref.~\onlinecite{watson01}) or 
CaV$_2$O$_5$ (Ref.~\onlinecite{konst00}) fall in this range of parameters. \\\indent
Further, we have discussed the density of states in the presence of a finite 
concentration of impurities in the limit of $J_{\mathrm{L}}\ll J_{\mathrm{R}}$ both numerically and analytically.
The comparison of the different approaches shows that diagrammatic methods 
give quantitatively correct results for small impurity concentrations and, furthermore, a qualitatively 
correct picture is obtained for large impurity concentrations. 
As the diagrammatic approaches neglect the interference of impurities and the effect 
of impurity clusters, we have presented a careful analysis of systems with small impurity clusters which allows
us to understand details visible in the density of states. 
Natural extensions of this work, i.e., the computation of observables and the discussion of
systems with arbitrary ratios of $J_{\mathrm{L}}/J_{\mathrm{R}}$, are left for future work. Nevertheless,
our results already imply the appearance of additional features in the spin gap which could be observed by, e.g., optical experiments on 
bond-disordered spin ladder materials.

\indent {\bf Acknowledgments - }
This work was supported by the DFG, Schwer\-punkt\-programm 1073, by
a DAAD-ANTORCHAS exchange program, and by CONICET and Fundaci\'on Antorchas, Argentina. We acknowledge support by the Rechenzentrum of the TU Braunschweig
where
parts of the numerical computations have been performed on a COMPAQ ES45 (CFGAUSS).


\end{document}